\documentstyle[11pt,epsf]{article}

\catcode`@=11
\long\def\@caption#1[#2]#3{\par\addcontentsline{\csname
  ext@#1\endcsname}{#1}{\protect\numberline{\csname
  the#1\endcsname}{\ignorespaces #2}}\begingroup
    \small
    \@parboxrestore
    \@makecaption{\csname fnum@#1\endcsname}{\ignorespaces #3}\par
  \endgroup}
\catcode`@=12
\newcommand{\newc}{\newcommand}
\newc{\fdiffg}{{\partial^2F_\gamma \over \partial E\partial \Omega}}
\newc{\fdiffx}{{\partial^2F_{\chi\chi\to\gamma\gamma} \over \partial E\partial
\Omega}}
\newc{\fomegag}{{\partial F_\gamma\over \partial\Omega}}
\newc{\fomegax}{{\partial F_{\chi\chi\to\gamma\gamma}\over \partial\Omega}}
\newc{\fomegageps}{{\partial F^\epsilon_\gamma\over \partial\Omega}}
\newc{\infinity}{\infty}
\newc{\vrel}{v_{\rm rel}}
\newc{\vcm}{v_{\rm cm}}
\newc{\acm} {a_{\rm cm}}
\newc{\bcm} {b_{\rm cm}}
\newc{\ccm} {c_{\rm cm}}
\newc{\arel} {a_{\rm rel}}
\newc{\brel} {b_{\rm rel}}
\newc{\crel} {c_{\rm rel}}
\newc{\gsim}{\lower.7ex\hbox{$\;\stackrel{\textstyle>}{\sim}\;$}}
\newc{\lsim}{\lower.7ex\hbox{$\;\stackrel{\textstyle<}{\sim}\;$}}
\newc{\mev}{{\rm\,MeV}}
\newc{\gev}{{\rm\,GeV}}
\newc{\tev}{{\rm\,TeV}}
\def\NPB#1#2#3{Nucl. Phys. {\bf B#1} (19#2) #3}
\def\PLB#1#2#3{Phys. Lett. {\bf B#1} (19#2) #3}
\def\PLBold#1#2#3{Phys. Lett. {\bf#1B} (19#2) #3}
\def\PRD#1#2#3{Phys. Rev. {\bf D#1} (19#2) #3}
\def\PRL#1#2#3{Phys. Rev. Lett. {\bf#1} (19#2) #3}

\def\ZPC#1#2#3{Zeit. f\"ur Physik {\bf C#1} (19#2) #3}

\def\MPL#1#2#3{Mod. Phys. Lett. {\bf#1} (19#2) #3}
\def\beq{\begin{equation}}
\def\eeq{\end{equation}}
\def\bea{\begin{eqnarray}}
\def\eea{\end{eqnarray}}

\begin{document}
\setlength{\baselineskip}{0.2in}

\begin{titlepage}

\begin{flushright}
{\large
UM-TH-94-38\\
November 1994\\
}
\end{flushright}
\vskip 2cm
\begin{center}
{\Large\bf
THEORY, PHENOMENOLOGY, AND PROSPECTS FOR \\
DETECTION OF \\
SUPERSYMMETRIC DARK MATTER
}
\vskip 1cm
{\Large
E.~Diehl\footnote{E-mail: {\tt ediehl@hep.uchicago.edu}}
\\}
\vskip 2pt
{\large\it Enrico Fermi Institute, University of Chicago, Chicago, IL 60637}\\
\vskip 0.2truein
{\Large
G.L. Kane\footnote{E-mail: {\tt gkane@umich.edu}},
Chris Kolda\footnote{E-mail: {\tt kolda@umich.edu}},
and
James D. Wells\footnote{E-mail: {\tt jwells@umich.edu}}
\\}
\vskip 2pt
{\large\it Randall Laboratory, University of Michigan, Ann Arbor,
MI 48109, USA}\\
\end{center}
\vskip .5cm
\begin{abstract}
One of the great attractions of minimal super-unified
supersymmetric models is the {\it prediction} of a massive, stable,
weakly interacting particle (the lightest supersymmetric partner, LSP)
which can have the right relic abundance
to be a cold dark matter candidate.  In this paper
we investigate the identity, mass, and properties of the LSP
after requiring gauge coupling unification,
proper electroweak symmetry breaking,
and numerous phenomenological
constraints.  We then discuss the prospects for detecting
the LSP.
The experiments which we investigate are
(1) space annihilations
into positrons, anti-protons, and gamma rays,
(2) large underground arrays
to detect upward going muons arising
from LSP capture and annihilation in the sun and earth,
(3) elastic collisions on
matter in a table top apparatus, and
(4) production of LSPs or decays into
LSPs at high energy colliders.
Our conclusions are
that space annihilation experiments and large underground detectors
are of limited help in initially detecting the LSP although perhaps
they could
provide confirmation of a signal seen in other experiments, while table
top detectors have considerable discover potential.
Colliders
such as LEP II,
an upgraded Fermilab, and LHC, might be the best dark matter detectors
of all.  This paper improves on most previous analyses in the literature
by (a) only considering parameters not already excluded by several physics
constraints listed above,
(b) presenting results that are independent of
(usually untenable) parameter choices, (c) comparing opportunities to study
the same cold dark matter, and (d) including minor technical improvements.

\end{abstract}
\end{titlepage}

\vfill\eject

\setcounter{footnote}{0}

\section{Introduction}

Recently there has been
a great deal of effort investigating minimal supergravity
models with
gauge coupling unification,
and radiatively generated
electroweak symmetry breaking (i.e., models in which the Higgs mechanism
is derived rather than being
imposed)~\cite{roberts1,arnowitt1,kelley1,lopez2,kane1,castano1,barger1,
carena1}.
It is quite remarkable that these minimal super-unified models
have survived the numerous fusillades by theorists and experimentalists
trying to constrain or even rule out supersymmetry.  We are then left
with the inescapable phenomenological
conclusion that minimal super-unified models
are at least as good as the Standard Model in describing
the world around us.  ``Super-unified'' means that the three gauginos are
given a common mass at the unification scale, and similarly the sfermions
and Higgses are all given a common mass there.

In addition to passing all current phenomenological tests, these supersymmetric
models have the potential to answer such vexing questions as, where does
all the dark matter come from?  It is remarkable that when all known
phenomenological constraints are applied
to the minimal super-unified solutions, those which remain often have
an LSP (Lightest Supersymmetric Particle)
which does not overclose the universe and which
generally has a significant enough relic density to be cosmologically
interesting -- that is, provide enough mass density to the universe to
be a viable cold dark matter candidate.

In this paper we primarily investigate the
LSP in the context of this constrained minimal supersymmetric standard
model~\cite{kane1} (CMSSM), and determine what predictions can be
made for its detection in many different
experiments.  The experiments we consider here are neutrino yield from
capture and annihilation of the LSP in the sun and earth, direct detection
methods with various nuclei, collider signatures, and space annihilation
signatures such as anomalous positron fraction, $\bar p$, and $\gamma$
ray production.

Although there is so far no clear phenomenological necessity to investigate
supersymmetric models which drastically differ from the minimal
super-unified one, we nevertheless do consider implications (such
as non-universal soft masses) when the issue becomes interesting.
But our main goal is to make a definitive statement on the properties and
detectability of the LSP within the minimal constrained super-unified model.

\section{The particle physics model}
\bigskip

Before continuing further, we define what
we mean by CMSSM (See Ref.~\cite{kane1}
for a complete description).  Briefly, CMSSM is the parameter space
defined by minimal particle
content (i.e. the Standard Model
spectrum with two Higgs doublets plus superpartners), the gauge group
$SU(3)\times SU(2)\times U(1)$ below the scale of gauge coupling unfication,
common gaugino masses, common scalar masses,
common trilinear and bilinear soft masses at the unification scale,
correct electroweak symmetry breaking
and conserved R--parity.
Futhermore, a CMSSM solution must satisfy
all known experimental constraints including the requirement that relic
particles not overclose the universe ($\Omega_{\rm TOT}h^2<1$).
We do not impose additional constraints such as proton decay limits
since that would require
more specific knowledge of the exact high scale theory than we believe
is available today.  In this regard, our analysis is more general than
previous ones~\cite{arnowitt1,lopez1}.

R--parity is extremely important to the analysis in this paper.
R--parity is a discrete $Z_2$ symmetry which, when conserved,
forbids baryon number
and lepton number violating interactions in the superpotential,
and dictates an absolutely stable LSP.
But, the most
general renormalizable and gauge invariant supersymmetric Lagrangian
manifestly breaks R--parity and leads to the prediction of unacceptably
rapid proton decay.  Therefore,
R--parity conservation was
hypothesized~\cite{farrar1,weinberg1,sakai1,dimopolous1} and
tentatively accepted
as the theoretical reason for a stable proton.
{}From the perspective of supersymmetrizing the standard model Lagrangian
one could argue that R--parity should not be viewed as an {\it ad hoc}
symmetry since the standard model
Lagrangian has no relevant B or L violating interactions.  Despite this,
few were pleased with ``old R--parity'' for several reasons:  (1) No one
knew where it came from, (2) it was a global symmetry and therefore quantum
gravity effects might possibly preclude its conservation, and (3) it was not
a unique option -- just forbidding B {\it or} L violating
terms is sufficient to save the proton.

Progress has been made over the last several years and a ``new R--parity''
can now be
viewed as more attractive.  Krauss and Wilczek~\cite{krauss1}
originally pointed out that discrete symmetries could result from spontaneous
symmetry breaking of a continuous gauge symmetry.  With a gauge symmetry
producing the low energy discrete symmetry, all ruminations
about the ills of global symmetries become irrelevant.  Furthermore, with this
possibility at our disposal we begin to understand that it might not
be difficult to motivate the very existence of a discrete symmetry like
R--parity from spontaneously broken gauge symmetries at the GUT scale.
Indeed, several authors~\cite{font1,martin1} have shown that R--parity
can be derived from a continuous
$B-L$ gauge symmetry, and that a large class of models
which contain $U(1)_{B-L}$ at the high scale automatically conserves
R--parity in the low energy phase.

The ``new R--parity'' which is much more highly motivated and understood
than the ``old R--parity'' comes with a price:  gauged $B-L$ at the high
scale.  There are several ways to envision a $B-L$ gauged symmetry.  We could
assume that it is an admixture of $U(1)$'s left over from compactification in
a string theory along with the Standard Model gauge group.  Or we can assume
that $B-L$ is a subgroup of a simple grand unified group (which may or may
not have come
from the string), in which case we acknowledge $SO(10)$ as the
leading candidate.  Certainly there are other possibilities, but our viewpoint
at the moment is that gauge coupling unification and gauged R--parity are
mutually compatible and perhaps even imply each other.  We expect R--parity
conservation to eventually
arise from the structure of the Higgs potential of the full
theory, but we do not consider its origin further in this paper.

Another important aspect of the analysis is radiative electroweak
symmetry breaking~\cite{ibanez2,inoue1,gaume1,ellis1},
wherein the Higgs mechanism,
and therefore symmetry breaking, can
be {\it derived} through the renormalization of the Higgs masses from
the high scale, $m^2_{H}(Q^2=M^2_{\rm GUT})>0$, to the low scale,
$m^2_{H}(Q^2=M^2_Z)<0$.  With a large Yukawa coupling graciously
provided by the top quark, such an aesthetic explanation of electroweak
symmetry breaking is manifest in the minimal super-unified model.
That the Higgs mechanism emerges from such a theory is one of its
most attractive features.
After GUT scale parameters are chosen and after all renormalization
group equations are solved, we must carefully analyze the
Higgs potential to see if it truly admits EWSB and we must make sure
that the minimum of the potential will numerically reproduce the
Z mass (i.e., $v_1^2+v_2^2=v^2=4M^2_Z/(g_1^2+g_2^2)$).  The relationship
among the low energy Lagrangian parameters required to recover the Z mass is
\begin{equation}
\label{ewsb}
{1\over 2} M_Z^2+\mu^2={m^2_{H_d}-m^2_{H_u}\tan^2\beta\over\tan^2\beta-1}
+L(\tan\beta,m_t,m_{\tilde t_i},\ldots)
\end{equation}
where $m_{H_u}$ and $m_{H_d}$ are the Higgs soft masses and
$L$ contains all effects of one loop corrections to the
Higgs potential.

We want to emphasize that proper electroweak symmetry breaking
(EWSB) is not a matter of taste or aesthetics; it is a
requirement imposed by nature.  Therefore, all analyses which do not
require Eq.~\ref{ewsb} be satisfied must necessarily suffer from
unphysical model solutions and suspect conclusions; they cannot correctly
describe our world.

To illustrate how important it is to require that the Higgs
potential admits electroweak symmetry breaking, we have constructed
a set of solutions which have EWSB imposed, and others which do not.
To illustrate, we have chosen
$m_0=300\gev$, $\tan\beta=3$,
$A_0=0$, $m_t=170\gev$, and ${\rm sgn}(\mu)=+$.
(Recall that $m_0$ is the universal high scale mass for all scalars,
$\tan\beta$ is the ratio of the vacuum expectation values of the up-Higgs
to the down-Higgs, $A_0$ is the universal tri-linear soft supersymmetry
breaking term at the high scale, and $\mu$ is the
Higgs mixing coefficient in the superpotential~\footnote{For definitions
of SUSY parameters and conventions used in this paper see Ref.~\cite{kane1}}.)
The actual numbers for
these parameters are not important since the results that we present
below are general for any set of input parameters.  We then varied
$M_2$ (or equivalently, $m_{1/2}$) and determined $\mu$ from the
requirement of EWSB.  The solid line in Fig.~\ref{figM2muz} shows where
%%%%%%%%%%%%%%%%%%%%%%%%%%%%%%%%%%%%%%%%%%%%%%%%%%%%%%%%%%%%%%%%%%%%%
\begin{figure}
\centering
\epsfysize=3in
\hspace*{0in}
\caption{Scatter plot of unphysical solutions in the $M_2$--$\mu (m_Z)$
plane with $m_0=300\gev$, $\tan\beta=3$, $A_0=0$, and $m_t=170\gev$ held
fixed.  These solutions do not satisfy the constraints of electroweak
symmetry breaking and yield an incorrect Z-boson mass.
The solid line, however,
is physical in that the Z-boson mass is fixed to its experimental value.}
\label{figM2muz}
\end{figure}
%%%%%%%%%%%%%%%%%%%%%%%%%%%%%%%%%%%%%%%%%%%%%%%%%%%%%%%%%%%%%%%%%%%%%
these solutions, which do have correct electroweak symmetry breaking,
lie in the $M_2$--$\mu$ plane.  We then made a second
set of solutions by choosing $M_2$ and $\mu$ randomly with values up
to 1 TeV.  Unlike the first set of solutions with EWSB imposed, the values
of $M_2$ and $\mu$ in the second set
are not correlated with each other (no EWSB) and
are represented by the dots in Fig.~\ref{figM2muz}.

We should emphasize again that the dots in Fig.~\ref{figM2muz} represent
unphysical supersymmetric models; only the line corresponds to a theory
with EWSB and the correct value for $M_Z$.
Although the solid line in Fig.~\ref{figM2muz} will move around with different
choices of $m_0$, $\tan\beta$, etc, the strong correlation between $M_2$
and $\mu$ will always exist, and needs to be calculated.
Unfortunately, one often finds studies of supersymmetry
phenomenology which effectively map the whole unphysical $M_2$-$\mu$ plane
(or some optimistic part)
onto
some particular observable (e.g., neutrino flux from neutralino capture and
annihilation in the sun).  We illustrate with one example
why such an approach can lead to orders of magnitude errors.

The example that we give is the spin dependent elastic cross ($\sigma_{SD}$)
section with LSPs impinging on ${}^{73}{\rm Ge}$.  Fig.~\ref{figsd}
%%%%%%%%%%%%%%%%%%%%%%%%%%%%%%%%%%%%%%%%%%%%%%%%%%%%%%%%%%%%%%%%%%%%%
\begin{figure}
\centering
\epsfysize=3in
\hspace*{0in}
\caption{A direct mapping of the solutions in Fig.~1, showing spin-dependent
cross sections on ${}^{73}{\rm Ge}$ vs. LSP mass.  The dots are
unphysical solutions which do not have proper electroweak symmetry breaking,
and the solid line is physical.}
\label{figsd}
\end{figure}
%%%%%%%%%%%%%%%%%%%%%%%%%%%%%%%%%%%%%%%%%%%%%%%%%%%%%%%%%%%%%%%%%%%%%
is a complete mapping of all the solutions of Fig.~\ref{figM2muz}
in the $M_2$--$\mu$ plane onto the $\sigma_{SD}$--$m_\chi$ plane.
Notice that the line (physical solutions) is
at least three orders of magnitude below the highest cross sections of
the dots (unphysical solutions), and that well over half of the dots
are above the line.  Since the counting rate for a nuclear detector
scales linearly with $\sigma_{SD}$ one could be three orders of magnitude
too optimistic about the prospects of detecting LSP cold dark matter
if the unphysical solutions are used, as has happened.

As discussed above, requiring the Higgs potential to admit EWSB is an
absolute requirement on any physically viable supersymmetric theory.
Furthermore, experimental predictions using supersymmetric solutions
which do not impose this requirement will often lead to grossly optimistic
conclusions.

\section{What is the LSP?}
\bigskip

Besides such exotics as axions and axinos, minimal supergravity
with conserved R--parity has several
stable dark matter possiblities.
If supersymmetry breaking is communicated to the visible sector through
gravitational interactions then the gravitino mass ($m_{3/2}$) is
of order the superpartner masses or greater.  Since we want
the superpartner masses to be
of order the weak scale to solve the naturalness problem
then the gravitino mass also must be that heavy.
Furthermore, decays of the gravitino with a weak scale mass
could greatly disrupt the
successful description of big bang nucleosynthesis~\cite{weinberg2}.
To escape this problem
the gravitino is usually assumed to be heavy enough not to affect
this analysis.

Ruling out all charged and colored objects~\cite{ellis4}, we are left with
two possibilities for the LSP:  the sneutrino and lightest neutralino.
The sneutrino as LSP was first suggested by the authors of
Ref.~\cite{hagelin2}
and was recently considered again in finer detail by the
authors of Ref.~\cite{falk1}.   One of the conclusions of Ref.~\cite{falk1}
was that
sneutrinos with cosmologically interesting relic densities were already
ruled out by experiment except for possibly a small region of parameter
space near $m_{\tilde \nu}\simeq 600\gev$.

We would like to point out, however, that within the minimal super-unified
approach the sneutrino cannot be the LSP and have a mass above $M_W$.
To demonstrate this we first write down the neutralino mass
matrix~\cite{haber1}
in the $\{\tilde B,\tilde W^3,\tilde H_d,\tilde H_u\}$ basis,
\beq
Y=\pmatrix{M_1 & 0 & -m_Z\sin\theta_W\cos\beta & m_Z\sin\theta_W\sin\beta \cr
           0 & M_2 & m_Z\cos\theta_W\cos\beta & -m_Z\cos\theta_W\sin\beta \cr
           -m_Z\sin\theta_W\cos\beta & m_Z\cos\theta_W\cos\beta & 0 & \mu \cr
           m_Z\sin\theta_W\sin\beta & -m_Z\cos\theta_W\sin\beta & \mu & 0 \cr},
\eeq
and note that an upper bound can be placed on
the lightest neutralino from the diagonal elements of the neutralino
mass matrix squared:
\begin{equation}
m^2_\chi\leq {\rm min}\{ (YY^\dagger)_{11},(YY^\dagger)_{22},
(YY^\dagger)_{33},(YY^\dagger)_{44} \}.
\end{equation}
Choosing just one of these,
\begin{equation}
m^2_\chi\leq (YY^\dagger)_{11}=M^2_1+M^2_Z\sin^2\theta_W
\end{equation}
we obtain a working upper bound on the LSP mass:
\begin{equation}
m^2_{\chi,{\rm max}}=M^2_1+M^2_Z\sin^2\theta_W=
     \eta m^2_{1/2}+M^2_Z\sin^2\theta_W
\end{equation}
where
\begin{equation}
\eta=\left( \frac{5}{3}\frac{\alpha_Y}{\alpha_{\rm GUT}} \right)^2
       \approx 0.18.
\end{equation}
The renormalized low scale sneutrino mass is
\begin{equation}
m^2_{\tilde \nu}=m^2_0+b_{\tilde \nu} m^2_{1/2}+{1\over 2}M^2_Z\cos 2\beta
\end{equation}
where $b_{\tilde \nu}\approx 0.5$~\cite{kane1}.  Therefore,
\begin{equation}
m^2_{\tilde \nu,{\rm min}}\approx {1\over 2} (m^2_{1/2}-m^2_Z).
\end{equation}
Since $m^2_{\tilde \nu}$ scales faster with $m_{1/2}$ than does $m^2_{\chi}$,
there must be a maximum value of $m_{\tilde \nu}$ for which it
is still possible for the sneutrino to be the LSP:
$m^2_{\chi, {\rm max}}=m^2_{\tilde \nu,{\rm min}}$.  This value is
a little less than $m_W$. Therefore, a sneutrino with mass greater than
$m_W$ cannot be the LSP, a result which is independent of all
parameters ($\tan\beta$, $\mu$, $m_0$, $\ldots$).
Within the CMSSM
we find no solutions with the sneutrino as LSP.  That is, imposing all
the physics constraints always leads to the LSP being the lightest neutralino.

Somewhat higher mass sneutrinos
as LSP could be obtained by relaxing universal gaugino mass conditions,
but one would be hard pressed to get $m_{\tilde\nu}\approx 600\gev$.
It is conceivable that such massive sneutrino LSPs are possible in
unusual extended gauge group theories, but we know of no examples.
We merely state that within
the present framework we must reject the sneutrino as a cosmologically
interesting LSP.  If it so happens that the sneutrino is the LSP with mass
less than $m_W$ then it will be detected at LEP II.  Sneutrinos in this
mass range do not give substantial $\Omega h^2$, so they are not interesting
cold dark matter candidates.

We are now left with the neutralino as the sole surviving dark matter
candidate.  Whereas a sneutrino with mass of ${\cal O}(m_W)$ yields
negligible and uninteresting relic abundance, the neutralino can yield
just the right amount of dark matter depending on its
composition~\cite{goldberg1,krauss2,ellis4}.
We express the LSP composition as a superposition of the four
supersymmetric weak eigenstates,
\beq
\chi=Z_{11}\tilde B+Z_{12}\tilde W^3+Z_{13}\tilde H_d+Z_{14}\tilde H_u
\eeq
A compelling argument for the minimal super-unified approach to
supersymmetry model building is that it automatically outputs a
neutralino as the LSP with the correct composition to be a cold
dark matter candidate.  Specifically, gaugino mass renormalization
from the high scale
and the requirement of radiative electroweak symmetry breaking invariably
produce a lightest neutralino which is Bino-like
(see Fig.~\ref{binorun} and caption).
%%%%%%%%%%%%%%%%%%%%%%%%%%%%%%%%%%%%%%%%%%%%%%%%%%%%%%%%%%%%%%%%%%%%%
\begin{figure}
\centering
\epsfysize=3.5in
\hspace*{0in}
\caption{We show how the renormalization group running of the gaugino mass
terms and $\mu$ automatically generate a bino-like LSP.  The lower solid
lines are $M_2$ (upper line) and $M_1$ (lower line).  Their initial values
are input at the high scale with the universal gaugino mass assumption.
The dotted lines
are the values of $m^2_{H_u}$ (lower line) and
$m^2_{H_d}$ (upper line).
(Note that $m^2_{H_u}$ runs negative at low scale as required by radiative
electroweak symmetry breaking.) Their values are input at the high scale
with the common scalar mass assumption.  The value of
$\mu$, here denoted by the dot-dashed line, is determined at the
low scale by the radiative electroweak
symmetry breaking condition of Eq.~\protect\ref{ewsb}, and the values at
other scales are calculated from the renormalization group equations.
The Bino ($\tilde B$) mass $M_1$ is the lightest mass parameter
in the neutralino mass matrix and so the LSP is a Bino-like neutralino.}
\label{binorun}
\end{figure}
%%%%%%%%%%%%%%%%%%%%%%%%%%%%%%%%%%%%%%%%%%%%%%%%%%%%%%%%%%%%%%%%%%%%%
This preferred requirement on the
neutralino wave function was first noticed outside the context of
supergravity model building~\cite{roszkowski1}.

The existence of this stable weakly interacting massive particle (WIMP)
in supersymmetry does not depend on astrophysics.
It is a property of a class of supersymmetry theories
which were motivated by reasons unrelated to what one might see in
a telescope.  Nevertheless, we do not believe that it is an accident
that supersymmetry predicts a WIMP with generically large relic
abundance, AND that the astrophysics data probably does imply the existence
of a WIMP with large relic
abundance~\cite{trimble1,primack2,strauss1,nusser1,white1,ostriker1}.
This position
is further strengthened by the paucity of visible matter compared to
inflation's preferred value of $\Omega_{TOT}=1$.
We do not wish to review further
the arguments for the existence of dark matter since our purpose is to
study what SUSY predicts regardless of the astrophysics and cosmology.
When needed, however, we do assume that the
local density of the dark matter in the
galactic halo is
$0.2~{\rm GeV/cm^3}<\rho_{loc}<0.4~{\rm GeV/cm^3}$~\cite{flores1}.
Unless otherwise stated, we alway use $\rho_{loc}=0.4~{\rm GeV/cm^3}$
in our numerical work.  Astrophysics and cosmology only enter our analysis
in the calculation of $\Omega_\chi h^2$ and in considering the
detection of LSPs, not in determining the LSP properties.

To obtain the relic density of the LSP, we calculate
using well known analytic
techniques~\cite{hut1,lee1,vysotskii1,olive1,srednicki1,wells2,roszkowski2}
and compare
with or use many formulas in the
literature~\cite{olive2,griest3,lopez1,macdonald1,ellis6,drees2}.
For each solution we calculate the thermal average by considering
all possible final states that the neutralino can
annihilate into, $f\bar f$, $W^+W^-$, $W^\pm H^\mp$, $ZZ$, $Zh$, $ZH$, $ZA$,
$hA$, $HA$, $hH$, $hh$, $HH$, $AA$, $H^\pm H^\mp$.  Furthermore, for each
solution we calculate the freeze-out temperature, photon reheating
factor, and the number of degrees of freedom at freeze-out in our attempt
to perform a precise relic density calculation.
We realize that the
analytic techniques used to solve the Boltzmann equation,
along with cosmological uncertainties,
can lead to a predicted relic abundance which is off by
20\% or more~\cite{kamionkowski1,griest1,gondolo1}.
One such difficulty, first pointed
out in Ref.~\cite{kane2} and quantitatively studied in
Ref.~\cite{gondolo1,griest1,nath2}, is the
possibility of neutralinos annihilating through a
resonance.  The Taylor series expansion of the annihilation cross-section
becomes untrustworthy and special techniques must be used to obtain
the correct relic density.
It is for these reasons that we use the relic density
calculation only as a rejection criteria and not as a measure of the
local density as some authors have done.  To be precise,
we reject all solutions which are not in the range expressed by
$0.05<\Omega_\chi h^2 <1.0$, and then assume that
$\rho_{loc}\approx 0.4~{\rm GeV/cm^3}$
independent of the calculated $\Omega_\chi h^2$.
It turns out that our results do not depend sensitively
on the above-quoted
lower and upper values of $\Omega h^2$ in our definition
for cosmologically interesting solutions as long as these values are
reasonable.  We have compared our relic density calculation with
others\footnote{We would like to thank M.~Drees and M.~Nojiri, and also
L.~Roszkowski,
for allowing us to compare results of our relic density calculations
with results from their programs.},
and have found good agreement.
Figure~\ref{omegah2} is a histogram of $\Omega_\chi h^2$
%%%%%%%%%%%%%%%%%%%%%%%%%%%%%%%%%%%%%%%%%%%%%%%%%%%%%%%%%%%%%%%%%%%%%
\begin{figure}
\centering
\epsfysize=3in
\hspace*{0in}
\caption{Histogram of $\Omega_\chi h^2$ for all solutions with $m_t=170\gev$.
Note the peak centered
at $\Omega_\chi h^2\approx 0.1$.  Given the measure on our
input parameter space and $h=0.35$, the
constrained solutions seem to prefer $\Omega_{CDM}\approx 0.8$ in the
``cosmologically
interesting'' region of $0.05<\Omega_\chi h^2<1$.}
\label{omegah2}
\end{figure}
%%%%%%%%%%%%%%%%%%%%%%%%%%%%%%%%%%%%%%%%%%%%%%%%%%%%%%%%%%%%%%%%%%%%%
for all solutions with $m_t=170\gev$.
We keep the solutions with $0.05<\Omega_\chi h^2<1$,
leaving us with
many solutions to analyze.

Before describing different techniques to detect the LSP we want to emphasize
one more time that each solution which we consider in the subsequent analysis
is ``fully constrained''.  That is, each solution is consistent with gauge
coupling unification, electroweak symmetry breaking, invisible width of the
Z, $b\to s\gamma$, all direct production collider constraints, etc.
Many past analysis have ignored important
phenomenological constraints in their models.
For example, Bottino {et al.}~\cite{bottino1,bottino2} consider a
large region of parameter space which is incompatible with the Z width
constraint and the $b\to s\gamma$ constraint.
(Note that ignoring the Z-width and
$b\to s\gamma$ constraints and allowing such low generic Higgs and
sfermion masses lead to substantially more optimistic conclusions for
LSP detectability~\cite{borzumati1}.)
Furthermore, they choose to drastically simplify their
SUSY paramater space by allowing many particles to obtain masses which
violate present experimental limits:
$m_h=50\gev$, $m_{\rm sfermions}=1.2m_\chi$
for all $m_\chi>45\gev$, etc.
We do not reduce the effective dimensionality of SUSY parameter space in such
an arbitrary manner, but rather choose a well motivated theory in
which to work.
We attempt to cover {\it all} of the {\it physically
realizable} supersymmetric parameter space, and allow only those parameters
which are consistent with the above-described theoretical and experimental
requirements.  In Fig.~\ref{spectrum}
%%%%%%%%%%%%%%%%%%%%%%%%%%%%%%%%%%%%%%%%%%%%%%%%%%%%%%%%%%%%%%%%%%%%%
\begin{figure}
\centering
\epsfysize=3.5in
\hspace*{0in}
\caption{A spectrum scatter plot of the constrained minimal supersymmetry
parameter space.  Each dot represents a mass of a particle particle labelled
directly below and belongs to a supersymmetric solution which satisfies
all theoretical and experimental requirements described in the text.
The horizontal banding is due to numerical sampling and is not
of significance.  The LSP is the first of the four vertical bands
collectively labeled $\chi^0$.}
\label{spectrum}
\end{figure}
%%%%%%%%%%%%%%%%%%%%%%%%%%%%%%%%%%%%%%%%%%%%%%%%%%%%%%%%%%%%%%%%%%%%%
we show a scatter plot of all masses in the CMSSM.  Each point represents
a mass of one particle in a supersymmetric solution which satisfies all
the above theoretical and experimental constraints.  In Fig.~\ref{spectrum1}
%%%%%%%%%%%%%%%%%%%%%%%%%%%%%%%%%%%%%%%%%%%%%%%%%%%%%%%%%%%%%%%%%%%%%
\begin{figure}
\centering
\epsfysize=3in
\hspace*{0in}
\caption{Mass spectra of a typical constrained minimal supersymmetry solution.
This model satisfies all the theoretical and experimental constrains
discussed in the text.  The LSP is the lightest of the four $\chi^0$
states.}
\label{spectrum1}
\end{figure}
%%%%%%%%%%%%%%%%%%%%%%%%%%%%%%%%%%%%%%%%%%%%%%%%%%%%%%%%%%%%%%%%%%%%%
we show just one solution out of the thousands contained in Fig.~\ref{spectrum}
to give the reader a feel for the typical mass correlations.
We now consider several different LSP detection schemes using the
CMSSM parameter space as shown in Fig.~\ref{spectrum} to
study LSP detection.

\section{LSP Detection through Space Annihilation}
\bigskip

While gently floating around in the galactic halo, the LSPs have some finite
probability of annihilating with other LSPs into quarks, leptons, gauge
bosons or Higgs bosons, which in turn fragment into
photons, positrons, and anti-protons
in the process.  These annihilation products then can find their way into
an earth-based or space-based detector, and the
measurement of anomalously large
fluxes of photons, positrons, or anti-protons could indicate a high density
of LSPs in the halo.  In this section, we present a quantitative measure
for the prospects of finding these space annihilation signals.
Note that all cases depend on rather uncertain astrophysics assumptions
and background estimates.
Before launching into the specifics of each method,
we first construct a
reference formalism which is applicable for all the methods.  Then we
will apply the formalism to each detection scheme independentently.

First, for annihilations into general two body final states,
$\chi\chi\to AB$, we define the ``annihilation strength function'' as
\beq
\label{asf}
S_{AB}(\vec r)={\rho^2(\vec r) \over m^2_{\chi}}(\sigma v)_{AB},
\eeq
($\rho(\vec r)$ is the mass density at position $\vec r$ and $v$ is
relative velocity)
which is interpreted as
\beq
[S_{AB}(\vec r)]={ {\rm \#~of~AB~pairs}\over {\rm cm^3~s}}~
{\rm at~position~}\vec r.
\eeq
Since the LSPs are highly non-relativistic in the
halo, the center of mass energy is $2m_\chi$.

But we are most often interested
in a particular annihilation product (say, particle $y$)
of the $AB$ pair, and so we must introduce a
general ``fragmentation yield function'' $F_{y/AB}(E,m_\chi)$
which describes the number and
energy distribution of the particle $y$ originating from the
decays of the $AB$ annihilation products.  The ``source strength function''
of the $y$ particles is then
\beq
\label{ssy}
{dS_y(E,\vec r)\over dE}=S_{AB}(\vec r) F_{y/AB}(E,m_\chi).
\eeq
By integrating Eq.~\ref{ssy} over energy one can easily see that
\beq
\int dE F_{y/AB}(E,m_\chi)={\rm ~\# ~of~}y{\rm 's}~{\rm per~1~}AB~
{\rm pair~produced~at~}E_{cm}/2=m_\chi.
\eeq

To obtain the area flux at a position $\vec r_{det}$ (a detector location)
we must propagate the particles from the position $\vec r$ to $\vec r_{det}$.
Furthermore, the energies of the produced particles could be
dramatically attenuated during its long journey to earth.
The effects of this
modulation will be described by a general ``transport modulation function''
$M_y(E,E')$.  In general $M_y$ does depend on position, but
we employ the standard
approximation and ignore this dependence.
Therefore, placing the detector at $\vec r=0$, defining $\theta=0$ as the
direction to the galactic center, and summing over all possible direct
annihilation products, we obtain a general formula for
the differential flux at the
detector:
\beq
{dA_y(E,\theta)\over dE d\Omega}=\sum_{AB} {1\over 4\pi}\int dE' \int dr
M_y(E,E') F_{y/AB}(E',m_\chi) S_{AB}(r,\theta)
\eeq
or
\beq
\label{fflux}
{dA_y(E,\theta)\over dE d\Omega}=\sum_{AB} {1\over 4\pi}\int dE'
M_y(E,E') F_{y/AB} (E',m_\chi) {(\sigma v)_{AB}\over m^2_{\chi}}
\int_{\rm halo} dr \rho^2(r,\theta).
\eeq
The integral over $r$ is required to sum the volume density production
rate at all points in the halo.
The $\theta$ dependence comes about from the coordinate transformation on
$\rho(\vec r)$ (the origin is now shifted from
the galactic center to the detector and therefore angular invariance of
the dark matter distribution is no longer preserved).
The units of $dA_y/dEd\Omega$ are
${\rm cm^{-2}~sec^{-1}~GeV^{-1}~sr^{-1}}$.

\subsection{Gamma Rays}
\bigskip

Since photons with energies in the GeV range
experience very little modulation in the galactic halo,
we can approximate $M_{\gamma}$ by $M_{\gamma}(E,E')=\delta(E-E')$.
Putting this into Eq.~\ref{fflux} gives us a working equation to analyze
the photon flux from LSP annihilation:
\beq
{dA_\gamma(E,\theta)\over dE d\Omega}=\sum_{AB} {1\over 4\pi}
F_{\gamma/AB}(E,m_\chi) {(\sigma v)_{AB}\over m^2_{\chi}}
\int_{\rm halo} dr \rho^2(r,\theta).
\eeq
The integral over $\rho^2$ can be represented by $r_{gc}\rho^2_{loc} I(\theta)$
where $r_{gc}$ is distance to the galactic center,
and $I(\theta)$ is an angular factor~\cite{gunn1,stecker1,turner1}
which depends on
the specifics of the halo distribution and is a maximum at $\theta =0$
(toward the galactic center).  Since we are most interested in diffuse
gammas from high galactic latitudes, we set $I(\theta )=1$~\cite{gunn1}.
When the LSPs directly annihilate into photons then
$F_{\gamma/AB}=2\delta(E-m_\chi)$ and the differential energy flux is
simply
\beq
{dA_\gamma\over dEd\Omega}={1\over 2\pi} \delta(E-m_\chi)
{\rho^2_{loc}\over m^2_\chi}r_{gc} (\sigma v)_{\gamma\gamma}.
\eeq

The diffuse photon background is not known for $E\gsim 1\gev$.
Using data from the $\mev$ region and extropolating to higher energies,
the photon spectrum is inferred to be~\cite{dermer1}
\beq
{d\bar A_\gamma\over dEd\Omega}=8\times 10^{-7}\left(
{1\gev\over E}\right)^{2.7} {\rm cm^{-2}~sec^{-1}~\gev^{-1}~sr^{-1}}.
\eeq
The bar on top of the $\bar A_\gamma$ indicates expected background flux.

In this section we investigate the possibility that neutralinos annihilating
in the galactic halo produce copious amounts of photons directly
($\chi\chi \to \gamma\gamma$) or indirectly
($\chi\chi\to AB\to\gamma {\rm 's}+X$)
which clearly stand out from the expected background.  First,
we compare the expected background flux to monochromatic photon flux
from direct neutralino annihilation.  A useful observable for this
comparison is the ratio of the integrated signal flux to the integrated
background flux:
\beq
R_\gamma=
\frac{\int_{m_\chi-\epsilon}^{\infty}
   \left( {dA_\gamma\over dEd\Omega} \right) dE}
   {\int_{m_\chi-\epsilon}^{\infty}
\left( {d\bar A_\gamma\over dEd\Omega} \right) dE}\approx
(1.6\times 10^{10})\left( {(\sigma v)_{\gamma\gamma}\over \gev^{-2}} \right)
\left( {\gev\over m_\chi} \right)^{0.3}.
\eeq

We calculate the direct neutralino annihilation rate into
photons~\cite{srednicki2,rudaz1,giudice2,bergstrom1,bouquet1,rudaz2}
for all solutions.  We do not consider the Higgs and chargino
loop contributions~\cite{bergstrom2,jungman2}
since they are only large for Higgsino-like LSPs which we do not have,
and which do not give cosmologically interesting $\Omega_\chi h^2$.
In Fig.~\ref{Rgamma}
%%%%%%%%%%%%%%%%%%%%%%%%%%%%%%%%%%%%%%%%%%%%%%%%%%%%%%%%%%%%%%%%%%%%%
\begin{figure}
\centering
\epsfysize=3in
\hspace*{0in}
\caption{Scatter plot of $R_\gamma$ vs. $m_\chi$ for all allowed
solutions.  For a solution to be detectable $R_\gamma \gsim 1$ is
required.  Near
the top mass some solutions could be detected with a high resolution
detector, capable of measuring the integrated gamma ray flux at
energies very close to and above the LSP mass.}
\label{Rgamma}
\end{figure}
%%%%%%%%%%%%%%%%%%%%%%%%%%%%%%%%%%%%%%%%%%%%%%%%%%%%%%%%%%%%%%%%%%%%%
we make a scatter plot of $R_\gamma$ vs. $m_\chi$ for all solutions.
(Each point represents a complete solution.)  We find that near
$m_\chi\approx m_t$ the signal to background ratio approaches 1,
and there is a possibility of detecting the neutralino through
this technique.  This enhancement near top threshold is due to an enhanced
top loop contribution in the coupling to the photons and the fact that
the gamma ray background is expected to be a
steeply falling function of energy.

A continuous spectrum of photons is also expected from LSP annihilations
and we have decayed all the final states of neutralino annihilations,
counted the photons, and analyzed their energy spectra to determine
whether or not a detectable signal is possible.
We use JETSET~\cite{sjostrand1} to determine the photon yield from
final state annihilation products. It appears that
generally the continuous photon spectrum from LSP annihilation is
not enough to clearly stand out from background. Fig.~\ref{contphoton}
%%%%%%%%%%%%%%%%%%%%%%%%%%%%%%%%%%%%%%%%%%%%%%%%%%%%%%%%%%%%%%%%%%%%%
\begin{figure}
\centering
\epsfysize=3in
\hspace*{0in}
\caption{Typical continuous photon spectrum from LSP annihilations.  The
dotted line is the expected background in the multi-GeV region, and the solid
line is the calculated photon spectrum from the decays of all final state
annihilation channels.  If this background estimate is correct, it
is difficult to detect the signal.}
\label{contphoton}
\end{figure}
%%%%%%%%%%%%%%%%%%%%%%%%%%%%%%%%%%%%%%%%%%%%%%%%%%%%%%%%%%%%%%%%%%%%%
represents a typical case where only a small portion of the signal
spectrum smoothly rises above the expected background, not enough
to be easily discernible.  Not only is the signal low, but the measurement
is extremely difficult even with a large signal because electron
contamination makes it difficult to identify legitimate diffuse
gamma rays.  However, we note that a signal does exist, so a
clever way to separate signal from background could be fruitful.
Many experiments, such as those conducted at the
Whipple observatory, only look at high energy gammas from a point
source and so partially can solve the problem of electron contamination by
employing a highly restrictive solid angle cut.  Such techniques
to enhance the electron/photon rejection ratio are not permissible
when trying to detect high energy diffuse gamma radiation.  Therefore, not
only is the experiment a very difficult one, but the signal is
low for LSP annihilations into photons.

\subsection{Anomalous Positron Fraction}
\bigskip

Just as photons can be created in LSP annihilations, making their way
to earth, so too can positrons.
The working equation for the production and propagation of
positrons~\cite{tylka1,turner2} is
not as simple as that for photons.  Additional complications arise since
positrons get ``trapped'' in the
galactic magnetic field and since their energy distributions broaden
and soften over time from Thomson scattering and synchrotron radiation.

A containment time $\tau_{e^+}$ is usually introduced to parametrize the
amount of time that a positron is trapped in the galactic halo.
Since $c\tau_{e^+}$ is estimated
to be several orders of magnitude beyond the characteristic length scale
of the galactic halo, the positrons in some sense sample the entire halo
before finally escaping.  This effect will raise the expected
flux at the earth and can be approximated by the following substitution
\beq
\int_{\rm halo} dr \rho^2(r,\theta) \rightarrow (c\tau_{e^+})\bar \rho^2
\eeq
where $\bar\rho^2$ is the average density over the galactic halo (or more
precisely, the containment volume).  As an approximation to this we
will identify $\bar\rho^2$ as the local density $\rho^2_{loc}$.
We have essentially replaced the length scale of the galactic halo with
the higher
length scale of the containment time.
Then using the model of Ref.~\cite{turner2} we can extract
$M_{e^+}(E,E')$:
\beq
M_{e^+}(E,E')={1\over E^2\tau_{e^+}} \exp\left[ {90\over \tau_{e^+}}
\left( {1\over E'}-{1\over E} \right) \right].
\eeq
The differential energy flux is
\beq
{dA_{e^+}\over dEd\Omega}=\sum_{AB} {1\over 4\pi} \int dE' M_{e^+}(E,E')
F_{e^+/AB}(E',m_\chi)(\sigma v)_{AB}
{\rho^2_{loc}\over m^2_\chi} (c\tau_{e^+}).
\eeq

Kamionkowski and Turner~\cite{kamionkowski3} have suggested that neutralino
annihilation into W's and Z's could produce an almost monochromatic
line spectrum of positrons from the decays of these vector bosons.
In order for the signal to be detectable in the $e^+/(e^-+e^+)$ spectrum,
the neutralino must be almost completely Higgsino-like in order to give
a sufficiently large annihilation cross section. And even
then, the effect is quite small and required the authors to multiply
their signal by a factor of 10 just for it to show up on their graphs.
Unfortunately, such Higgsino-like LSPs do not give cosmologically
interesting $\Omega_\chi h^2$ nor do they occur in the CMSSM spectrum,
and the signal which we obtain from positrons via vector
boson production and decays is even smaller than that obtained by
Kamionkowski and Turner.  In Fig.~\ref{Rlinepos} we plot
%%%%%%%%%%%%%%%%%%%%%%%%%%%%%%%%%%%%%%%%%%%%%%%%%%%%%%%%%%%%%%%%%%%%%
\begin{figure}
\centering
\epsfysize=3in
\hspace*{0in}
\caption{Scatter plot of $R^{VV}_{e^+}$ vs. $m_\chi$ for all solutions with
$m_t=170\gev$.  Only for values of $R^{VV}_{e^+}\gsim 1$ would we expect
a significant enough bump in the $e^+/{(e^-+e^+)}$ spectrum to indicate
possible LSP annihilations through vector boson pairs. Our LSPs,
therefore, are not detectable through this mode.}
\label{Rlinepos}
\end{figure}
%%%%%%%%%%%%%%%%%%%%%%%%%%%%%%%%%%%%%%%%%%%%%%%%%%%%%%%%%%%%%%%%%%%%%
$R^{VV}_{e^+}$ vs. $m_\chi$ where $R^{VV}_{e^+}$ is defined
to be
\beq
R^{VV}_{e^+}={e^+~{\rm from~}\chi\chi\to W^+W^-,ZZ\over e^+~{\rm total}}~~~~
{\rm at}~E=E_{\rm peak}.
\eeq
Clearly, positrons from vector boson annihilation final states
cannot provide a clear signature of LSP annihilations in the galactic halo.

We can extend the analysis and sum over all final states of the LSP
annihilations and then determine the number and energy of final state
positrons and electrons.  The relevant $F_{e^+/AB}$ are
determined by the JETSET Monte Carlo~\cite{sjostrand1}.
One final state of particular
interest is the top quark which decays approximately 100 percent of the time
to $bW^+$ and so produces a positron from the subsequent $W^+$ decay.
Unfortunately, even this mode ($\chi\chi \to t\bar t$) will not produce
a significant enough positron source to make a large bump in the
$e^+/(e^-+e^+)$ spectrum.  In Fig.~\ref{bestep}
%%%%%%%%%%%%%%%%%%%%%%%%%%%%%%%%%%%%%%%%%%%%%%%%%%%%%%%%%%%%%%%%%%%%%
\begin{figure}
\centering
\epsfysize=3in
\hspace*{0in}
\caption{Our best case scenario is plotted for the case of $m_\chi=275\gev$
using
all positrons (and electrons) from the decays of all annihilation final
states (not just vector bosons).
Here the result is somewhat encouraging, although the signal to
background is still small.  However, a signal does exist, so it is
worth looking
for a better way to reduce background.}
\label{bestep}
\end{figure}
%%%%%%%%%%%%%%%%%%%%%%%%%%%%%%%%%%%%%%%%%%%%%%%%%%%%%%%%%%%%%%%%%%%%%
we plot the best case scenario which we find among our solutions.  For
this solution $m_\chi=275\gev$ and $\Omega_\chi h^2 =0.05$.  The two humps come
from top quarks decaying into positrons, and bottom quarks and tau's decaying
into positrons.  Note that if we were to scale the LSP density with
the calculated $\Omega_\chi$ we would have obtained a substantially
smaller signal.  We therefore conclude that positrons from LSP annihilations
will be a difficult signal to extract above the background but perhaps
not impossible.

\subsection{Anti-protons}
\bigskip

At low energies the ratio of anti-protons to protons in
the cosmic ray spectrum is expected~\cite{protheroe1} to be extremely low
(less than about $10^{-5}$ for $E<1\gev$).
When the LSP annihilates into quarks and gluons,
the stable particles in the final state after hadronization will in general
contain protons and anti-protons.  If enough anti-protons are created
and they hit the detector at low energies then this
would be a possible signal of LSP annihilations~\cite{silk1,hagelin1}.

Since the proton is trapped in the galactic magnetic field and has a rather
large containment time compared to the size of the galaxy, we can use
a very similar formalism to that which was used for positrons.  Namely,
we rewrite the integral over $r$ in Eq.~\ref{fflux} as
\beq
\int_{\rm halo} dr \rho^2(\vec r)\rightarrow
(v_{\bar p}\tau_{\bar p}) \rho^2_{loc}
\eeq
and then the differential energy flux becomes
\beq
{dA_{\bar p}\over dE d\Omega}=\sum_{AB} {1\over 4\pi}\int dE'
M_{\bar p}(E,E') F_{\bar p/AB}(E',m_\chi){\rho^2_{loc}\over m^2_\chi}
(\sigma v)_{AB} (v_{\bar p}\tau_{\bar p}).
\eeq
We use $\tau_{\bar p}=5\times 10^7~{\rm yrs}$ in our numerical calculations,
but the reader should be aware that there is at least a factor of ten
uncertainty in this number~\cite{ellis5}.

The fragmentation function $F_{\bar p/AB}(E',m_\chi)$ we extract from
JETSET.  The transport modulation function $M_{\bar p}(E,E')$ (here due
to solar wind effects) can
be extracted from Ref.~\cite{perko1} and is
\beq
M_{\bar p}(E,E')=
\delta(E'-f(E))\theta(p_c-p)+\delta(E'-E-\Delta E)\theta(p-p_c)
\eeq
where
\beq
f(E)=p_c{\rm ln} {p+E\over p_c+E_c}+E_c+\Delta E.
\eeq
Furthermore, we have chosen $p_c=1.015\gev$ and $\Delta E=0.5\gev$ which
corresponds to the minimum solar activity.

We present the results in Fig.~\ref{pbar} of our anti-proton signal analysis
as a scatter plot of $R_{\bar p}$ vs. $m_\chi$ where $R_{\bar p}$ is
defined as
\beq
R_{\bar p}={\rm \#~signal~\bar p{\rm 's}~with~100 \mev< E < 200\mev\over
\#~background~\bar p{\rm 's}~with~100\mev < E < 200 \mev}
\eeq
%%%%%%%%%%%%%%%%%%%%%%%%%%%%%%%%%%%%%%%%%%%%%%%%%%%%%%%%%%%%%%%%%%%%%
\begin{figure}
\centering
\epsfysize=3in
\hspace*{0in}
\caption{Scatter plot of $R_{\bar p}$ vs. $m_\chi$ for all solutions.
$R_{\bar p}$ is the ratio of low-energy antiprotons expected from LSP
annihilation to those expected from spallation in the energy range
of $100\mev$ to $200\mev$.
Detecting an effect
would probably require $R_{\bar p}$ to approach unity.
Detecting LSP annihilations
through an anomalously large anti-proton flux appears to be quite difficult
since the signal $\bar p$'s are never more numerous than the
expected background $\bar p$'s in this energy window.
Note that the region below
$m_\chi\approx 30\gev$ will be covered at FNAL and LEP (see section 7).}
\label{pbar}
\end{figure}
%%%%%%%%%%%%%%%%%%%%%%%%%%%%%%%%%%%%%%%%%%%%%%%%%%%%%%%%%%%%%%%%%%%%%
Our results are considerably less optimistic than those found in
Ref.~\cite{jungman1}.  However, the reasons are clear.  When radiative
symmetry breaking is included in the analysis there are correlations between
the LSP and squark masses, and the squarks are typically
heavier than what is usually optimistically
assumed otherwise.  Therefore the production
of gluons through squark-quark loops, which Ref.~\cite{jungman1} found to
be so important, become quite small and the number of anti-protons is
reduced.  We note again that there is a signal, and so as sensitivity
to $\bar p{\rm 's}$ is increased at low energies, a shelf in the
energy spectrum could become visible, indicating halo annihilations.

\section{LSP Detection through Earth and Sun Capture}
\bigskip

Given enough LSPs and a sufficient interaction strength between them and
ordinary matter, LSPs would
be captured by the Sun or Earth and
would annihilate
into neutrinos which could then be detected as upward going muons
in underground detectors.  Data on upward muons from Earth and
Sun from several underground experiments, notably
Kamionkande~\cite{kamiokande1}
and IMB~\cite{losecco1}, has been used to establish limits on
WIMPs~\cite{kamionkowski2}.
This technique will be exploited in the
future by several experiments such a MACRO
in Italy, the largest neutrino
telescope currently in operation, and several potentially much
larger underwater or under ice neutrino telescopes such as Dumand
off Hawaii, Amanda at
the South pole, Baikal in Lake Baikal, and Nestor off the Greek island
of Pylos.  The hope is that a $\rm 1~km^{2}$ neutrino telescope may
eventually be constructed which would greatly expand the potential of
this technique.  In this section we examine LSP detection by
this method.

\subsection{ Expected flux of upward muons from LSP annihilation }

The basic scenario for LSP detection via the observation of upward
muons in an underground detector is that halo LSPs lose energy
and are captured by
scattering from nuclei in either the Earth or Sun.  Subsequent
scatterings result in the LSPs settling to the center whereupon LSP
and anti-LSP annihilate into, among other things, neutrinos.  The
neutrinos travel relatively freely to the Earth, where they may
undergo a charged current interaction and produce an upward going muon
in an underground detector.  Only high energy ($>1\gev$)
neutrinos may produce upward going muons, and the only known high
energy neutrinos are those produced in atmospheric cosmic ray showers.
A sharp reduction in this background may be made by the considering
angular windows around the Earth and Sun.

We calculate the expected rate of upward
muons using the following steps (for more detail see~\cite{diehl1}):

\begin{enumerate}

\item Calculate the capture rate of LSPs by the Earth and Sun
due to scalar and spin-dependent scattering.
We use the formula of Ref.~\cite{gould1,kamionkowski2} for the capture rate
calculations.
The fraction of full capture rate~\cite{griest4} has been
considered taking into account the competition
between capture and annihilation rates.

\item Calculate the LSP annihilation branching ratios
into all relevant final states.

\item Determine the neutrino yield from all possible
annihilation products of the LSP.  We take into consideration
energy loss of heavy quark annihilation products~\cite{ritz1}.
We use JETSET~\cite{sjostrand1} to determine neutrino yield from the
subsequent decays of the annihilation products (e.g., $b\to\nu_e+X$).
We keep track of $\nu_{\tau}$ yield since it can undergo a charge-current
conversion to a $\tau$ which subsequently decays into a
$\mu$ with a branching fraction of $0.187$.

\item Determine energy loss and flux loss of neutrinos undergoing
neutral current or charged current interactions as they travel
through the sun~\cite{ritz1}.

\item Calculate the upward going muon rate given neutrino flux.
This is done by first determining the charged current conversion rate
in rock~\cite{morfin1}, and then by simulating the muon energy loss as it
propogates through the rock on its way into the detector~\cite{lohmann1}.

\item Finally, estimate the atmospheric neutrino
background~\cite{gaisser1,volkova1},
and scale it to recent experimental data~\cite{diehl1,kamiokande1}
for higher accuracy.

\end{enumerate}

It is a complicated question to evaluate whether the Earth or Sun will
yield a better upward muon signal.  The answer depends on LSP mass,
as well as the LSP interaction cross sections and annihilation
branching ratios.
[We have compared our spin-dependent and spin-independent cross section
calculations with those of Drees and
Nojiri
and find excellent agreement for
solutions with higher mass LSPs ($m_\chi\gsim 100\gev$).  For solutions
with lower mass LSPs we invariably have somewhat higher cross sections,
which leads to a larger signal of upward going muons
(and LSP-nucleon scattering which we discuss in the next section).]
Crudely speaking, for LSPs less than about 100 GeV
and having substantial spin-independent interactions, the Earth signal
could be larger.  For heavy LSPs and LSPs with primarily
spin-dependent interactions, the signal from the Sun will be larger.
In the case of CMSSM neutralinos considered here, the signal from the
Sun is almost always far larger than that of the Earth since
gaugino-like neutralinos have primarily spin-dependent interactions
for which the Earth provides little signal. Hence,
we show only the detection possibilities from upward muons
from LSP annihilations in the Sun.

We present our results in Fig.~\ref{SUNAREA} as the area required to
have an upward-going muon signal above background with a $4\sigma$
significance.
%%%%%%%%%%%%%%%%%%%%%%%%%%%%%%%%%%%%%%%%%%%%%%%%%%%%%%%%%%%%%%%%%%%%%
\begin{figure}
\centering
\epsfysize=3.5in
\hspace*{0in}
\caption{The detector area required for detection of upward going muons with
a $4\sigma$ statistical significance above background.  Such a signal would
be possible evidence for LSP capture and annihilation in the sun.
Only a few solutions with $m_\chi \leq 40\gev$ would be ruled out if
a $1~{\rm km^2}$ detector did not observe an effect, and those solutions
are detectable at LEP II and FNAL.}
\label{SUNAREA}
\end{figure}
%%%%%%%%%%%%%%%%%%%%%%%%%%%%%%%%%%%%%%%%%%%%%%%%%%%%%%%%%%%%%%%%%%%%%
In order to make this plot it was assumed that the
idealized detector is located at the equator (which has the largest
exposure to the sun), and is 100\% efficient in detecting upward muons
(i.e. any muon below the horizontal) of energy greater than 2 GeV.
The background of atmospheric upward muons was considered in the
angular bins required to observe 90\% of the LSP upward muon flux.
These bin sizes were determined
by a Monte Carlo performed by the Kamionkande collaboration
\cite{kammc}.  The detector area required
depends on the exposure time to the sun.  (Higher latitudes mean lower
exposure time). The areas in Fig.~\ref{SUNAREA}
must be multiplied by the factors listed in Table~\ref{AREAFRAC} to
account for this effect in actual experiments.  Also, we have not corrected
for different energy thresholds in different detectors, which would decrease
the effectiveness of Amanda and Dumand relative to the MACRO thresholds
we use.
%%%%%%%%%%%%%%%%%%%%%%%%%%%%%%%%%%%%%%%%%%%%%%%%%%%%%%%%%%%%%%%%%%%%%%%%
\table
\begin{center}
\begin{tabular}{|c|l|} \hline
Latitude   & Factor \\  \hline\hline
$0^{\circ}$            & 1.   \\ \hline
$20^{\circ}$  (DUMAND) & 1.09 \\ \hline
$42.5^{\circ}$ (MACRO) & 1.46 \\ \hline
$90^{\circ}$  (AMANDA) & 3.43 \\ \hline
\end{tabular}
\vspace{24pt}
\caption{Exposure factors for various detector locations.
The areas in Fig.~12 must be multiplied by the appropriate scale
factor.}
\label{AREAFRAC}
\end{center}
\endtable
%%%%%%%%%%%%%%%%%%%%%%%%%%%%%%%%%%%%%%%%%%%%%%%%%%%%%%%%%%%%%%%%%%%%%%

Our results for detectability of upward going muons are considerably
less optimistic than previous analyses~\cite{halzen1}.
The highly correlated mass spectra in minimal supergravity
(see Fig.~\ref{spectrum1}) forbid one from randomly
choosing squark masses, Higgs masses, gaugino masses, and LSP composition
independent from one another. The resulting parameter space then contains
gaugino-like LSPs with typically small spin-dependent cross-sections
and even smaller spin-independent cross-sections.
Furthermore,
our refinements in each step of the calculation (such as including
thresholds) turned out to reduce the signal even further.
Note that the largest working neutrino telescope, MACRO, has an area
of only about $\rm 10^{-3}~ km^{2}$ and so cannot constrain any
neutralino models with one year's observation.
Unfortunately, only a few solutions with $m_\chi \lsim 40\gev$ are
detectable even with a $1~{\rm km}^2$ neutrino telescope, and those
will be covered by LEP II and FNAL.

\section{LSP Detection through elastic collisions with nuclei}
\bigskip

LSPs can also be detected by directly searching for their elastic
scattering off nuclei in a table top detector.  There have been many
calculations of the LSP elastic scattering cross section off
matter~\cite{goodman1,kane2,ellis3,griest2,srednicki3,giudice1,drees1}
which we use.
As the illustration
of this technique, we have chosen to investigate two separate elements
which have different nuclear properties:
${}^{73}{\rm Ge}$, which has (essentially) just one unpaired neutron,
and ${}^{93}{\rm Nb}$, which has (essentially) just one unpaired proton.
Both these elements have large nuclear masses,
and therefore the coherent spin-independent
scattering amplitude, which couples to the
mass of the nucleus rather than the nucleon, will be enhanced.

Germanium is an excellent case study
because the nuclear properties
have been studied quite thoroughly~\cite{ressell1}
making the calculation of expected
rates more reliable than with other elements where one must rely on
simplistic models of nuclear properties.  Furthermore, a strong experimental
program is already underway~\cite{barnes1,shutt1}.
Although there is still some uncertainty coming from the quark spin content
of nucleons, these errors have diminished substantially
with more precise experimental numbers and improved theoretical
understanding of the measurements which have been conducted over
the years~\cite{ellis7,close1}.
Therefore, quark spin content of the
nucleons is no longer a major source of uncertainty in the spin dependent
cross section calculation.
We utilize the values for $\Delta q$ ($\Delta u=0.83\pm 0.03$,
$\Delta d= -0.43\pm 0.03$,
and $\Delta s= -0.10\pm 0.03$ for the proton) as suggested by
Ref.~\cite{ellis7}.

The current sensitivity of LSP-nucleon collisions is generally quoted
as 0.1 events per day, with future goals focussing on about 0.01 events
per day sensitivity.  With this in mind we plot in Figs.~\ref{Gemass}
and~\ref{Nbmass} the counting rate per kilogram per year for
a $\rm {}^{73}Ge$ and a $\rm {}^{93}Nb$ detector.
%%%%%%%%%%%%%%%%%%%%%%%%%%%%%%%%%%%%%%%%%%%%%%%%%%%%%%%%%%%%%%%%%%%%%
\begin{figure}
\centering
\epsfysize=3in
\hspace*{0in}
\caption{Scatter plot of the counting rate for the ${}^{73}{\rm Ge}$ nuclei
vs. $m_\chi$.  The upper line is the limit that would be placed on the rate
if $1~{\rm kg}$ of Ge were collecting data for 1 year and could suppress
background enough to be sensitive to $0.1~{\rm event}/{\rm kg}/{\rm day}$.
Likewise, the lower line is for $10~{\rm kg}$ of Ge with sensitivity
up to $0.01~{\rm events}/{\rm kg}/{\rm day}$ in which case many solutions
would be either detected or ruled out for $m_\chi\lsim 200\gev$.}
\label{Gemass}
\end{figure}
%%%%%%%%%%%%%%%%%%%%%%%%%%%%%%%%%%%%%%%%%%%%%%%%%%%%%%%%%%%%%%%%%%%%%
%%%%%%%%%%%%%%%%%%%%%%%%%%%%%%%%%%%%%%%%%%%%%%%%%%%%%%%%%%%%%%%%%%%%%
\begin{figure}
\centering
\epsfysize=3in
\hspace*{0in}
\caption{Scatter plot of the counting rate for the ${}^{93}{\rm Nb}$ nuclei
vs. $m_\chi$.  See Fig.~13 caption for further explanation.}
\label{Nbmass}
\end{figure}
%%%%%%%%%%%%%%%%%%%%%%%%%%%%%%%%%%%%%%%%%%%%%%%%%%%%%%%%%%%%%%%%%%%%%
We note that LSPs up to
approximately $100\gev$ are detectable with a $\rm 1~kg$ $\rm {}^{73}Ge$
detector.  Furthermore, if the sensitivity is increased by a factor of 10
(0.01 events per day) as some suggest is possible then solutions with
$m_\chi$ up to about $200\gev$ are detectable with $\rm 10~kg$.

In Fig.~\ref{compare} we compare the prospects of detecting the
%%%%%%%%%%%%%%%%%%%%%%%%%%%%%%%%%%%%%%%%%%%%%%%%%%%%%%%%%%%%%%%%%%%%%
\begin{figure}
\centering
\epsfysize=3.5in
\hspace*{0in}
\caption{Scatter plot comparing the effectiveness of a
neutrino telescope to a Ge table top experiment.  All solutions in the
lower left quadrant would either be detected by the neutrino telescope and/or
the table top experiment.  Note that $10~{\rm kg}$ of Ge covers many more
solutions than a $1~{\rm km^2}$ neutrino telescope.}
\label{compare}
\end{figure}
%%%%%%%%%%%%%%%%%%%%%%%%%%%%%%%%%%%%%%%%%%%%%%%%%%%%%%%%%%%%%%%%%%%%%
LSP using a $\rm 1~km^2$ neutrino telescope and a
10 kg $\rm {}^{73}Ge$ detector.
The lower left quadrant corresponds to the region where the LSP
will be detected by at least one of the methods.
The 10 kg Germanium detector covers a substantially larger fraction
of the parameter space than the $\rm 1~km^2$ neutrino telescope.
This leads us to conclude
that a 10 kg Germanium detector has more discovery potential than a
1~${\rm km}^2$ neutrino telescope.
But we should keep in mind that if table
top experiments are to have a significant constraining impact on super-unified
models the sensitivity must increase to 0.01 events per day {\it and} 10 kg or
more of nuclear detector material will be needed.

\section{LSP Detection through Collider Experiments}
\bigskip

If the LSP mass is less than $m_Z/2$ then experiments at the
$Z$--peak could possibly witness $Z\to \chi\chi$ decays.  Since
the neutralinos are invisible to the detectors of LEP and SLD,
we must ensure that such decays do not violate the invisible
width constraints.  The maximum contribution possible to the invisible
$Z$ width from non-standard physics
is $\Delta \Gamma\lsim 12~{\rm MeV}$~\cite{pdb94}.

Fig.~\ref{lspinv}
%%%%%%%%%%%%%%%%%%%%%%%%%%%%%%%%%%%%%%%%%%%%%%%%%%%%%%%%%%%%%%%%%%%%%
\begin{figure}
\centering
\epsfysize=3in
\hspace*{0in}
\caption{Constraints on the LSP mass and mixing
from the Z invisible width contributions of the
LSP and from the direct search limits on the lightest chargino.  The region
above the curved dotted line is excluded by $\Delta \Gamma_{inv}$ width
constraint.  The region to the left of the vertical dotted line is excluded
by the non-observation of $\chi^+\chi^-$ production at LEP.  Both of these
constraints must be applied on solutions with $m_\chi < m_Z/2$.  The dots
are solutions which pass these cuts.}
\label{lspinv}
\end{figure}
%%%%%%%%%%%%%%%%%%%%%%%%%%%%%%%%%%%%%%%%%%%%%%%%%%%%%%%%%%%%%%%%%%%%%
shows the constraints on the LSP from the invisible $Z$
width in the $\eta_Z$--$m_\chi$ plane where $\eta_Z$ is the neutralino
mixing factor in the coupling of the $Z$ with the LSP
($\eta_Z=|Z^2_{13}-Z^2_{14}|$).
An invisible width contour
is shown for $\Gamma_{\rm inv}=12\mev$.  All neutralinos which fall
to the left or above this curve are excluded from the invisible
width constraint.
There is also a constraint on the neutralino from non-observation
of $Z\to\chi^+\chi^-$ decays.  This constraint is possible because
of correlations between the chargino mass matrix and the neutralino
mass matrix through $\mu$, $M_2$ and (in the case of common $m_{1/2}$)
$M_1$.  We have used $\tan\beta >1.3$
to plot this chargino correlation
bound on the LSP, because
$\tan\beta<1.3$ would cause the top quark
Yukawa coupling to go strong below the high scale for $m_t\gsim 150\gev$
(which is implied by top searches at Fermilab~\cite{cdf1,d01} and
LEP precision data~\cite{novikov1,ellis2}), so the theory would no longer be
perturbative (as it is apparently observed to be by the gauge coupling
unification).
Since the minimal super-unified model almost invariably gives large
gaugino fraction, we expect that there will be many viable
solutions with small $\eta_Z$ and $m_\chi<m_Z/2$.  In Fig.~\ref{lspinv}
we have superimposed
a scatter plot of super-unified solutions to show that low mass $m_\chi$
are copious and do not violate the $\Gamma_{\rm inv}$ bound.

Measurements of $R_b=\Gamma (Z\to b\bar b)/\Gamma (Z\to {\rm had})$
at LEP might be indicating that the lightest stop and chargino are
less massive than about $m_W$~\cite{wells1}.  However, if universal
gaugino masses at the GUT scale are the correct description of nature,
then supersymmetric solutions which predict $R_b$ within one standard
deviation of the experimental measurement cannot provide sufficient
dark matter to be cosmologically interesting. This is because
scenarios with good $R_b$ require
the LSP
to be almost pure Higgsino-like, and annihilations through the $Z$ boson
become much too efficient and the final relic density is negligible.
By employing non-minimal GUT scale boundary conditions one could
``decouple'' the neutralino from the Z-boson (e.g., choose $M_1<< M_2$)
and allow for a large relic abundance of LSPs.  One has to worry
about squark and slepton $t$-channel exchanges yielding large cross
sections as well, but that is a problem uncommon scalar masses
can solve.  In any event, if supersymmetry does enhance $R_b$ to
within its $1\sigma$ experimental limit, then the lightest stop and/or
the lightest chargino will be discovered at LEP~II and the mass of the
LSP would be quickly measured.  Determining
whether or not the LSP is in fact the cold
dark matter would require other experiments which could directly
measure its existence in the halo, or experiments which would pin down
all other relevant parameters in the supersymmetric Lagrangian for a
reliable calculation of the expected relic abundance.

To a first approximation the mass reach at LEP
for charginos
and stops will be somewhere close to
half the beam energy.  Exceptions to this first approximation include
stops mixing such that the lightest stop does not couple to the $Z$,
or that the light stop and/or chargino is close enough to the LSP mass
that they decay into very soft leptons or jets and the events
are never detected.  Not much can be done about the first problem, which
reduces the sensitivity to stops by a few GeV (since there is
still a photon diagram), but
the latter problem can be overcome by making direct measurements of
the ``invisible rate'' from initial state photon radiation.   So if
superpartners are being created in large numbers
but then are decaying to untriggerable
final states, the invisible rate measurement should register it.

We have studied our solutions in an effort to determine the likelihood
of the above-mentioned detection scenarios.
In the case of the chargino there is always sufficient mass difference between
it and the LSP to provide the leptons or jets with plenty of energy to
make chargino production a clear signal at LEP II (provided the chargino
is light enough to be produced).  Then mass reconstruction algorithms
could conceivably determine the mass of the LSP from these events.
The same is true for many of the low mass stop solutions, but there
are some solutions which have a light stop and LSP which are nearly
degenerate.  For such solutions the stop events could not be directly tagged
and the LSP mass could not be reconstructed.
It is also possible, though not too likely, for
$\tilde \chi^+\rightarrow \tilde t_1 +b$, followed by
$\tilde t_1\to c+{\rm LSP}$, which would make reconstruction of the
LSP mass rather difficult.
It is interesting to note
that all the solutions which yield direct tagged chargino or stop events
at LEP II have an LSP mass below about $50\gev$.

A signal of supersymmetry also can be deduced cleanly at hadron
colliders from the production of charginos and
neutralinos~\cite{frere1,baer1,nath1,baer2}.
For a large region of parameter
space these particles will decay with a large branching fraction
into the LSP plus
three isolated, high $p_T$
leptons plus missing $E_T$. One such process is
represented by the Feynman diagram in Fig.~\ref{feynNC}.
%%%%%%%%%%%%%%%%%%%%%%%%%%%%%%%%%%%%%%%%%%%%%%%%%%%%%%%%%%%%%%%%%%%%%
\begin{figure}
\centering
\epsfxsize=3in
\hspace*{0in}
\caption{One example Feynman diagram representation of the supersymmetric
trilepton event.}
\label{feynNC}
\end{figure}
%%%%%%%%%%%%%%%%%%%%%%%%%%%%%%%%%%%%%%%%%%%%%%%%%%%%%%%%%%%%%%%%%%%%%
We simulate these trilepton signal events using a full Monte Carlo
event generator with a toy detector simulator.  To reduce background
we make the following cuts: one ``trigger'' lepton
with $|\eta|<1.0$ and $p_T>10\gev$; the two remaining leptons must
have $|\eta|<2.5$ and $p_T>4\gev$; $\Delta R_{ll}>0.4$; and
$m_{l^+l^-}\neq M_Z\pm 10\gev$.
The standard model
background rate is less than about $1~{\rm fb}$~\cite{baer2} given
similar background cuts and a sufficiently well-designed detector which
reduces lepton fakes.

Fig.~\ref{trilepton}
%%%%%%%%%%%%%%%%%%%%%%%%%%%%%%%%%%%%%%%%%%%%%%%%%%%%%%%%%%%%%%%%%%%%%
\begin{figure}
\centering
\epsfysize=3in
\hspace*{0in}
\caption{Scatter plot of expected trilepton event rates at Fermilab
for the $\chi^+_1\chi^0_2\to 3l$ signal with the cuts described in the text.
The upper line is the 10 event contour with $100~{\rm pb^{-1}}$
of data.  The lower dotted line is the 10 event contour with $1~{\rm fb^{-1}}$.
The mass reach for the latter is $m_\chi\gsim m_Z$ independent of all
parameters such as $\tan\beta$, etc.}
\label{trilepton}
\end{figure}
%%%%%%%%%%%%%%%%%%%%%%%%%%%%%%%%%%%%%%%%%%%%%%%%%%%%%%%%%%%%%%%%%%%%%
summarizes the results of our preliminary simulation~\cite{kane3}.
We make a scatter
plot of $m_\chi$ vs. $\epsilon_{\rm tot}\sigma(3l)$
for our entire parameter
space with $m_\chi$ below $100\gev$. (The variable $\sigma (3l)$ is the
total cross section for trilepton production through chargino plus
neutralino, and $\epsilon_{\rm tot}$
is the total efficiency from the previously mentioned detector cuts.)
Each point on
the graph represents a complete minimal super-unified model solution.
The top horizontal line is the 10 event contour with
$100~{\rm pb^{-1}}$, and the
lower horizontal line is the 10 event contour with $1~{\rm fb^{-1}}$ of
data.  From Fig.~\ref{trilepton} we see that with $1~{\rm fb^{-1}}$ of data,
all neutralinos below about $90\gev$ will be detected (indirectly)
through the trilepton signal.  This result is completely independent
of parameters in the model, such as $m_0$ and $\tan\beta$, etc.
Since the LSP occurs in these decays, its presence can be deduced
in principle from the
energy it carries away, and its mass and couplings inferred from an analysis
of the events.  Note that it is important to include the
solution-by-solution detectability since that can vary considerably
even for a given $m_\chi$.

At the LHC gluino production and decay through like-sign leptons or
jets with large missing $E_T$ might be the best way to detect
the neutralino.  Using different modes one can obtain viable signals of
gluino pair production for gluinos above $1~{\rm TeV}$~\cite{rubbia1}.
Since the gluino mass is strongly correlated with the lightest neutralino
mass ($m_\chi\sim 0.15 m_{\tilde g}$), the determination of the
gluino mass essentially determines the neutralino mass.  If a $1~{\rm TeV}$
gluino were discovered, then we would know that the LSP mass is approximately
$150\gev$, assuming the data were consistent with the CMSSM.

The above is a summary of some of the more promising ways to get at the
LSP mass at a collider.  But there are more ways.  Indeed, if R--parity
is conserved then a very model independent statement is that all particles
which reach the detector are standard model states plus an even number
of LSPs:
\beq
p\bar p\to 2n\chi+X_{\rm sm}.
\eeq
If a supersymmetry signal is found through {\it any} channel (like-sign
dileptons, tri-leptons, ...) then one always has a fighting chance at
reconstructing the LSP mass given enough events and sufficient understanding
of the theory.  Extracting results such as the LSP mass from LHC signals
may be very difficult.  To our knowledge the most likely technique may be
one using the structure of the theory, as described in section 9.7 of
Ref.~\cite{kane1}.
We suggest, therefore, that colliders be viewed as an
important player in dark matter detection. The mass reach of an upgraded
Tevatron and certainly the LHC rival and generally exceed other
experiments which have dark matter detection as one of their main goals.
Of course, what is explicitly demonstrated once superpartners are observed
at a collider is the existence of an LSP that lives longer than
about $10^{-8}$ seconds.  Direct confirmation of its role as dark matter
will {\it require} one or more of the previously discussed experiments.

\section{Conclusions and Comments}

We have carried out a comparative analysis of supersymmetric cold dark
matter and its detectability.  We use the constrained minimal supersymmetric
standard model (CMSSM) so all solutions examined are known to describe
nature, including gauge unification, electroweak symmetry breaking, and
experimental data.  As the experimental data increases and our knowledge
of the theory increases, the parameter space of CMSSM continues to shrink.
For each set of CMSSM parameters ($m_0$, $m_{1/2}$,
$\mu$, $A_0$, $\tan\beta$) a unique SUSY spectrum of masses
and couplings is calculated.  All LSP annihilation and scattering
diagrams and the resulting rates are calculated for those parameters.
Since all solutions satisfy all known constraints, we show
scatter plots of results for the allowed parameter sets rather than
selecting arbitrarily among them.
Our conclusions about detectability
do not confirm those of many previous analyses, because these earlier analyses
(a) did not satisfy conditions such as gauge coupling unification
or electroweak symmetry breaking,
(b) did not use constrained parameters from all experimental
data simultaneously, (c) chose optimistic and inconsistent sets
of parameters, or (d) made other optimistic assumptions.  We emphasize
that the scatter plots we have shown, while not eliminating some
dependence on the basic SUSY assumptions, do not depend at all on the
usual parameters ($\tan\beta$, $m_0$, $m_{1/2}$, etc.)  They are the best that
one can do without new (not generally available) sources of knowledge.

After constructing many supersymmetric solutions and then subsequently
using this same set of solutions to calculate
the LSP detection signal of various currently
running or proposed experiments,
we find lower detection prospects than were previously thought (except
in the case of dark matter detection by high energy colliders).
We note that in spite of large uncertainties in astrophysical
assumptions, which affect particularly the halo annihilation and solar
annihilation cases, the range of possible rates coming from lack of
knowledge of SUSY parameters is probably the dominant uncertainty in
calculating detectability. That means that doing the SUSY physics right
is the most important thing, and that is what we have emphasized.

\section*{Acknowledgements}
We would like to acknowledge G.~Tarl\'e and
S.~Martin for
useful discussions and suggestions regarding the manuscript.  We would
also like to thank S.~McKee, G.~Liu, and S.~Mikeyev for helpful
correspondences.  And special thanks go to
M.~Drees, M.~Nojiri, and L.~Roszkowski for graciously
allowing us to use the output of their computer programs to compare
with ours.  This work was supported in part by the U.S. Department of Energy.

\end{document}